 \documentclass[final,3p,times,twocolumn]{elsarticle}

\usepackage{amssymb}

\biboptions{sort&compress}

\journal{NIM B}

\begin{document}

\begin{frontmatter}



\title{Activation cross-sections of longer lived radioisotopes of deuteron induced nuclear reactions on terbium up to 50 MeV}


\author[1]{F. T\'ark\'anyi}
\author[1]{S. Tak\'acs}
\author[1]{F. Ditr\'oi\corref{*}}
\author[2]{A. Hermanne}
\author[4]{A.V. Ignatyuk}
\cortext[*]{Corresponding author: ditroi@atomki.hu}

\address[1]{Institute of Nuclear Research of the Hungarian Academy of Sciences (ATOMKI),  Debrecen, Hungary}
\address[2]{Cyclotron Laboratory, Vrije Universiteit Brussel (VUB), Brussels, Belgium}
\address[4]{Institute of Physics and Power Engineering (IPPE), Obninsk, Russia}

\begin{abstract}
Experimental cross-sections are presented for the first time for the $^{159}$Tb(d,xn)$^{155,157,159}$Dy, $^{155,156,160}$Tb and $^{153}$Gd nuclear reactions up to 50 MeV. The experimental data are compared with theoretical predictions of the ALICE, EMPIRE and TALYS nuclear reaction codes. Integral thick-target yields are also derived for the reaction products that have practical applications. 
\end{abstract}


\end{frontmatter}


\section{Introduction}
\label{1}
The aim of the work was manifold:
\begin{itemize}
\item	Very few experimental data exist in the literature for lanthanides, both for proton and deuteron induced reactions.
\item	The quality of the database for activation cross-sections of deuteron induced reactions is poor, compared to what exists for proton induced reaction. In our laboratories a systematic study of the activation cross-sections of deuteron induced reactions for different applications is in progress up to 50 MeV \cite{1,2}.
\item	Several radionuclides of the lanthanide group are becoming increasingly important in the field of nuclear medicine (diagnostic and therapeutic radioisotopes) \cite{3,4,5}.
\item	The prediction capability of the nuclear reaction model codes is still limited in case of deuteron induced reactions. Further improvement requires comparisons with experimental data. As terbium is monoisotopic, the targetry is simpler and cheaper and the cross-section results for a given activation product are not infected by contributions of multiple processes on the different stable target isotopes, hence no need for use of enriched target technology.
\item	 Activation cross-sections on Tb are required for production of the medically related $^{159}$Dy \cite{6} and $^{157}$Dy \cite{7,8,9} and need to be compared with other possible production routes.
\end{itemize}

In a literature search only two experimental cross-sections data sets were found, published by  Duc et al. \cite{10} for production of $^{159}$Dy, $^{157}$Dy and $^{160}$Tb up to 27 MeV and by Siri et al. \cite{11} for production of $^{157}$Dy and $^{160}$Tb up to 27 MeV. Mukhammedov  et al. \cite{12} measured integral yields of $^{159}$Dy as a function of the deuteron energy up to 12 MeV for charged particle activation analysis.

\section{Experimental and data evaluation}
\label{2}
The general characteristics and procedures for irradiation, activity assessment and data evaluation (including estimation of uncertainties) were similar to what is discussed in several earlier works of our group \cite{13,14}.
The main experimental parameters for the present study are summarized in Table 1. The principal methods used in data evaluation and the used decay data are collected in Table 2 \cite{13,15,16,17,18,19,20,21,22} and in Table 3. The good overlap of the re-measured excitation function for the $^{27}$Al(d,xn)$^{24}$Na monitor reaction with the recommended values \cite{23} is illustrated in Fig. 1.

\begin{table}[ht]
\tiny
\caption{Main experimental parameters for 2 irradiation sessions}
\centering
\begin{center}
\begin{tabular}{|p{0.75in}|p{1.in}|p{1.in}|} \hline 
\textbf{Reaction} & \textbf{${}^{159}$Tb(d,x)} & \textbf{${}^{159}$Tb(d,x)} \\ \hline 
 &  &  \\ \hline 
Incident particle & Deuteron & Deuteron  \\ \hline 
Method  & Stacked foil & Stacked foil \\ \hline 
Target and thickness  & Tb  foil, 95.47  $\mu$m & Tb  foil, 95.47  $\mu$m \\ \hline 
Number of Tb  target foils & 15 & 7 \\ \hline 
Target composition & Tb, Dy, Al & Tb, Eu, Sb, Pb, Al \\ \hline 
Accelerator & Cyclone 90 cyclotron of the Université Catholique in Louvain la Neuve (LLN) & Cyclone 90 cyclotron of the Université Catholique in Louvain la Neuve (LLN  \\ \hline 
Primary energy & 50 MeV & 50 MeV \\ \hline 
Covered energy range & 49.6-7.9 & 48.5-36.1 \\ \hline 
Irradiation time & 60 min & 93 min \\ \hline 
Beam current & 104.94 nA & 103.46  nA \\ \hline 
Monitor reaction \cite{23} & ${}^{27}$Al(d,x)${}^{24}$Na  reaction    & ${}^{27}$Al(d,x)${}^{24}$Na  reaction \\ \hline 
Monitor target and thickness & ${}^{nat}$Al,  26.76 $\mu$m & ${}^{nat}$Al,  49.06 $\mu$m \\ \hline 
detector & HpGe & HpGe \\ \hline 
$\gamma$-spectra measurements & 4 series & 4 series \\ \hline 
Cooling times after EOB & 4.3-7.3 h,\newline  49.8-69.5 h, \newline 581.7-652.5 h\newline 2277.0-2422.8 h& 7.0-8.3 h, \newline 45.1-47.2 h\newline 256.6-364.4 h,\newline 2373.3-2432.4 h\newline  \\ \hline 
\end{tabular}
\end{center}
\end{table}

\begin{table}[ht]
\tiny
\caption{Main parameters of data evaluation (with references)}
\centering
\begin{center}
\begin{tabular}{|p{1.in}|p{1.in}|p{0.6 in}|} \hline 
Gamma spectra evaluation & Genie 2000, Forgamma & \cite{15.16} \\ \hline 
Determination of beam intensity & Faraday cup (preliminary)\newline Fitted monitor reaction (final) &  \cite{17} \\ \hline 
Decay data & NUDAT 2.6 & \cite{18} \\ \hline 
Reaction Q-values & Q-value calculator & \cite{19} \\ \hline 
Determination of  beam energy & CityplaceAnderson (preliminary)\newline Fitted monitor reaction (final) & \cite{20}\newline \cite{13}  \\ \hline 
Uncertainty of energy & Cumulative effects of possible uncertainties &  \\ \hline 
Cross-sections & Reaction cross-section on element and on single isotope &  \\ \hline 
Uncertainty of cross-sections & Sum in quadrature of all individual linear  contributions & \cite{21} \\ \hline 
Yield & Physical yield & \cite{22} \\ \hline 
\end{tabular}
\end{center}
\end{table}

\begin{table*}[ht]
\tiny
\caption{Investigated radionuclides, their decay properties and production routes}
\centering
\begin{center}
\begin{tabular}{|p{0.6in}|p{0.5in}|p{0.6in}|p{0.5in}|p{0.7in}|p{0.8in}|} \hline 
Nuclide decay mode & Half-life & E$_\gamma$(keV) & I$_\gamma$(\%) & Contributing reaction & Q-value\newline (keV) \\ \hline 
\textbf{${}^{159}$Dy\newline }$\varepsilon $: 100 \%\textbf{} & 144.4 d & 58.0 & 2.27 & ${}^{159}$Tb(d,2n) & -3372.48 \\ \hline 
\textbf{${}^{157}$Dy\newline }$\varepsilon $: 100 \%~\textbf{} & 8.14 h & 182.424\newline 326.336 & 1.33\newline 93 & ${}^{159}$Tb(d,4n) & -19260.72 \\ \hline 
\textbf{${}^{155}$Dy\newline }$\varepsilon $: 100 \%~\textbf{} & 9.9 h & 184.564\newline 226.918 & 3.37 \%\newline 68.4 & ${}^{159}$Tb(d,6n) & -35671.3 \\ \hline 
\textbf{${}^{153}$Dy\newline }$\alpha $: 0.0094\newline $\varepsilon $: 99.9906\textbf{\newline } & 6.4 h & 80.723\newline 99.659\newline 213.754\newline 254.259\newline 274.673\newline 389.531\newline 537.225\newline 1023.99\newline  & 11.1\newline 10.51\newline 10.9\newline 8.6 \newline 3.1\newline 1.52\newline 1.33\newline 1.09 & ${}^{159}$Tb(d,8n) & -51824.12 \\ \hline 
\textbf{${}^{1}$${}^{60}$Tb\newline }$\beta $${}^{-}$: 100 \%\textbf{} & 72.3 d & 86.7877\newline 197.0341\newline 215.6452\newline 298.57 83\newline 879.378\newline 962.311\newline 966.166\newline 1177.954 & 13.2\newline 5.18\newline 4.02\newline 26.1\newline 30.1\newline 9.81\newline 25.1\newline 14.9 & ${}^{159}$Tb(d,p) & 4150.644 \\ \hline 
\textbf{${}^{1}$${}^{56}$Tb\newline }$\varepsilon $: 100 \% & 5.35 d & 88.97\newline 199.19\newline 262.54\newline 296.49\newline 356.38\newline 422.34\newline 534.29\newline 1065.11\newline 1154.07\newline 1159.03\newline 1222.44\newline 1421.67 & 18\newline ~41\newline 5.8\newline 4.5\newline 13.6\newline ~8.0\newline ~67\newline ~10.8\newline ~10.4\newline ~7.2\newline 31\newline 12.2 & ${}^{159}$Tb(d,p4n) & -25880.05 \\ \hline 
\textbf{${}^{1}$${}^{55}$Tb\newline }$\varepsilon $: 100 \% & 5.32 d & 86.55\newline 105.318\newline 148.64\newline 161.29\newline 163.28\newline 180.08\newline 340.67\newline 367.36 & 32.0\newline 25.1\newline 2.65\newline 2.76\newline 4.44\newline 7.5\newline 1.18\newline 1.48 & ${}^{159}$Tb(d,p5n)\newline ${}^{155}$Dy decay & -32794.5 \\ \hline 
\textbf{${}^{1}$${}^{53}$Gd\newline }$\varepsilon $: 100 \%\textbf{} & 240.4 d & 97.43100\newline 103.18012 & 29.0\newline 21.1 & ${}^{159}$Tb(d,2p6n)\newline $^{153}$Dy decay\newline  & -46519.36 \\ \hline 
\end{tabular}
\end{center}
\begin{flushleft}
\footnotesize{\noindent When complex particles are emitted instead of individual protons and neutrons the Q-values have to be decreased by the respective binding energies of the compound particles: np-d, +2.2 MeV; 2np-t, +8.48 MeV; n2p-${}^{3}$He, +7.72 MeV; 2n2p-$\alpha$, +28.30 MeV}
\end{flushleft}
\end{table*}

\begin{figure}[h]
\includegraphics[scale=0.3]{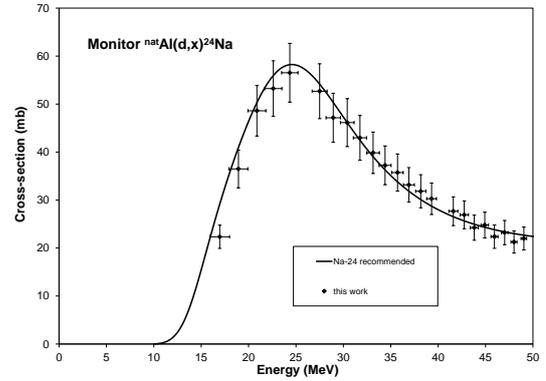}
\caption{The simultaneously measured monitor reactions for determination of deuteron beam energy and intensity}
\end{figure}

\section{Theoretical calculations}
\label{3}
We compared our measured cross-sections with predictions of the ALICE-IPPE \cite{24}, EMPIRE \cite{25} and TALYS \cite{26} nuclear reaction codes. The TALYS results were taken from the TENDL 2011 and TENDL 2012 nuclear reaction libraries \cite{27}. We present both versions to show the improvements and the still remaining problems between the two versions of the code. In case of ALICE and EMPIRE we have used the codes modified for better description of deuteron induced reactions (EMPIRE-D and ALICE-IPPE D) described in more detail in our previous reports \cite{28,29}.

\section{Results and discussion}
\label{4}

\subsection{Excitation functions}
\label{4.1}
The measured cross-sections for the production of the activation products are presented in Table 4 and Figs. 2-8. All cross-section values of dysprosium radionuclides are due to direct production via (d,xn) reactions. The terbium radioproducts are produced only directly via (d,pxn) reactions or additionally through the decay of the shorter-lived isobaric parent dysprosium radioisotope (cum). The investigated $^{153}$Gd is produced directly via (d,2pxn) reaction (including complex particle emission) and from the decay of parent $^{153}$Tb  radioisotope. The ground state of the produced radioisotopes  above the direct production can be produced additionally through the internal transition of the isomeric state. The cross-section is marked with (m+) when the half-life of the isomeric state is significantly shorter comparing to the half-life of the ground state and the cross-sections of the production of ground state were deduced from spectra after nearly complete decay of the isomeric state.  

\subsubsection{Production of $^{159}$Dy (T$_{1/2}$=144.4 d)}
\label{4.1.1}
According to Fig. 2 our data are systematically higher than the earlier  results of Duc et al. \cite{10}. The agreement with the results of the theoretical codes is acceptable, not counting the generally observed underestimation of the high energy cross-sections by the theory.

\begin{figure}[h]
\includegraphics[scale=0.3]{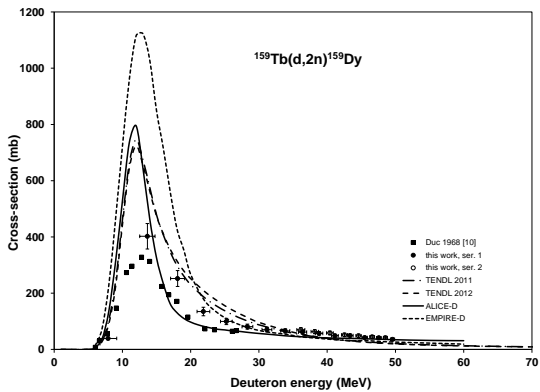}
\caption{Experimental and theoretical excitation functions of the $^{159}$Tb(d,2n)$^{159}$Dy  reaction}
\end{figure}

\subsubsection{Production of  $^{157}$Dy (T$_{1/2}$=8.14 h)}
\label{4.1.2}
The present and the earlier experimental data for the $^{159}$Tb(d,4n)$^{157}$Dy reaction are shown in Fig. 3. The agreement between all experimental data is acceptable. No explanation for the second bump around 50 MeV predicted by TENDL 2011 and 2012 was found. For comparison of the productions routes the cross-sections of Lebowitz et al. \cite{7} for the $^{159}$Tb(p,3n) reaction are also shown. 

\begin{figure}[h]
\includegraphics[scale=0.3]{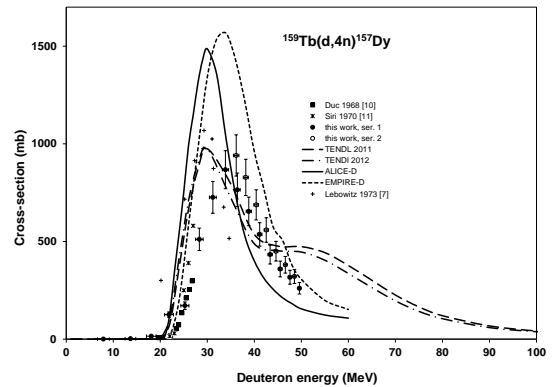}
\caption{Experimental and theoretical excitation functions of the $^{159}$Tb(d,4n)$^{157}$Dy  reaction}
\end{figure}

\subsubsection{Production of  $^{155}$Dy (T$_{1/2}$=9.9 h)}
\label{4.1.3}
No earlier experimental data were found in the literature. Comparison with the theory in the overlapping energy range shows a good agreement for EMPIRE-D, while the two other codes overestimate by a factor of 2 to 5 at 50 MeV (Fig. 4). No explanations, however for the second bump around 90 MeV predicted by TENDL 2011 and 2012.

\begin{figure}[h]
\includegraphics[scale=0.3]{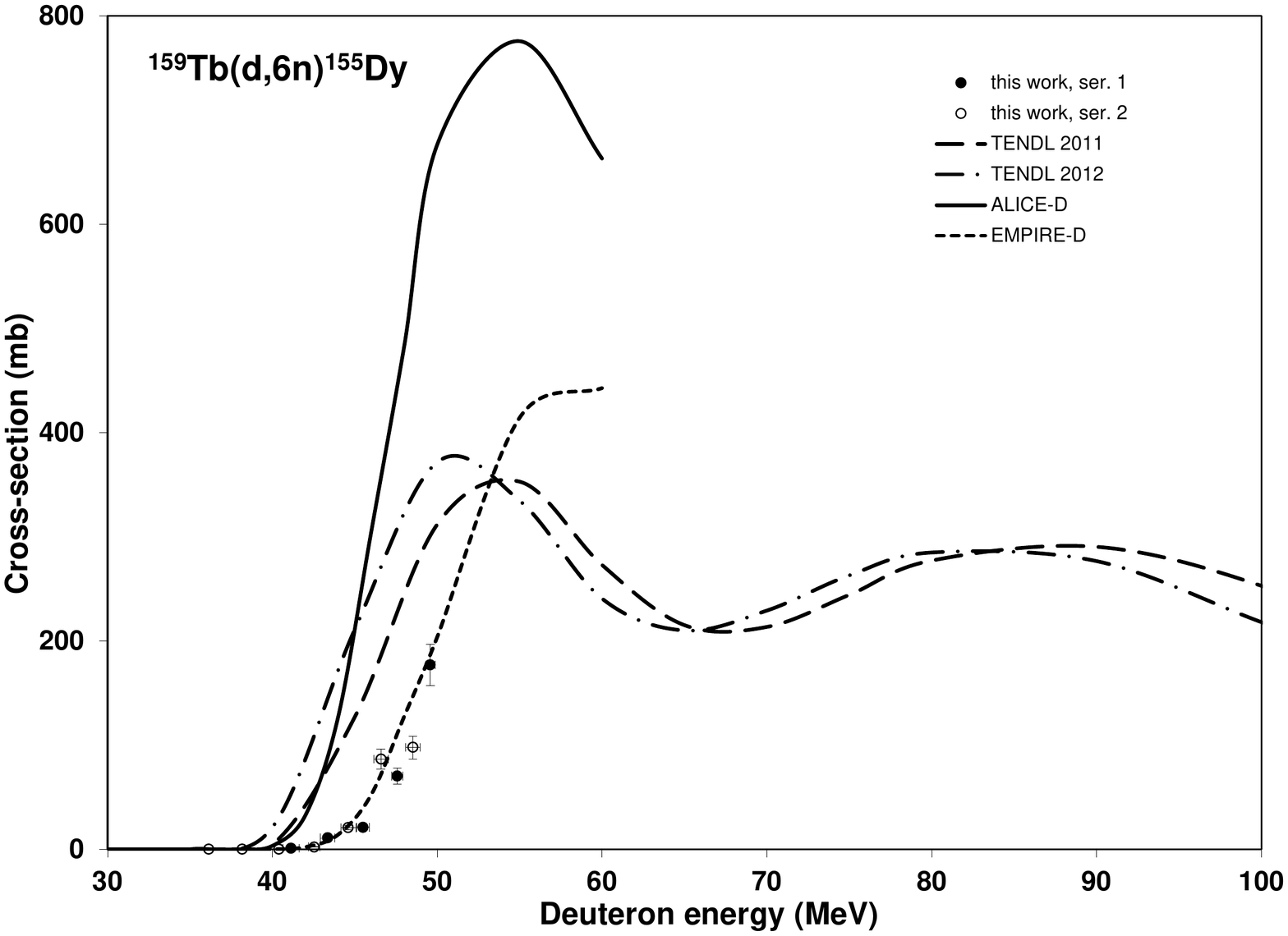}
\caption{Experimental and theoretical excitation functions of the $^{159}$Tb(d,6n)$^{155}$Dy  reaction}
\end{figure}

\subsubsection{Production of  $^{160}$Tb (T$_{1/2}$=72.3 d)}
\label{4.1.4}
We have only a few experimental data points at low energies around the maximum of the excitation function (Fig. 5). Our experimental data are significantly higher than the experimental data of Duc et al \cite{10} and a little lower compared to the data of Siri et al. \cite{11}. The theoretical predictions of the two TENDL libraries are a factor of two low, compared to the experimental data, while EMPIRE-D and ALICE-D reproduce rather well the shape and values of our experiment.

\begin{figure}[h]
\includegraphics[scale=0.3]{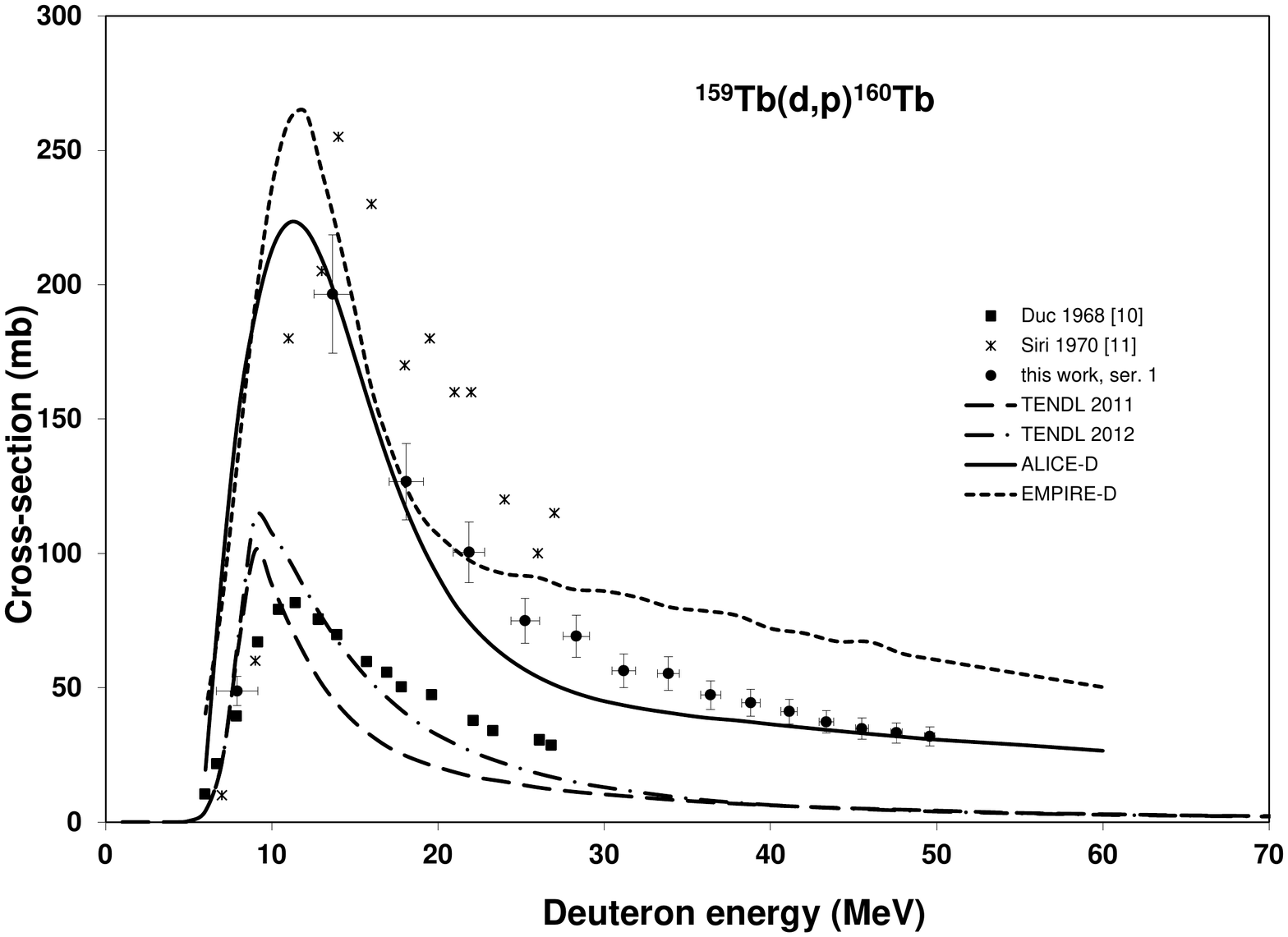}
\caption{Experimental and theoretical excitation functions of the $^{159}$Tb(d,p)$^{160}$Tb  reaction}
\end{figure}

\subsubsection{Production of  $^{156}$Tb  (T$_{1/2}$=5.35 d) (m+)}
\label{4.1.5}
The $^{156}$Tb is produced only directly via (d,p4n) reaction. The $^{156}$Tb has a long-lived ground state (T$_{1/2}$=5.35 d) and two shorter-lived isomeric states (5.3 h and 24.4 h) decaying completely to the ground state. Our experimental data for production of ground state (Fig. 6) were deduced from spectra after nearly complete decay of the isomeric states (m+). No earlier experimental data exist for comparison. The data in TENDL libraries show that the ground state is produced mostly directly. The contribution from the isomeric decay is small. In the overlapping energy range the theoretical data slightly overestimate the experimental results.

\begin{figure}[h]
\includegraphics[scale=0.3]{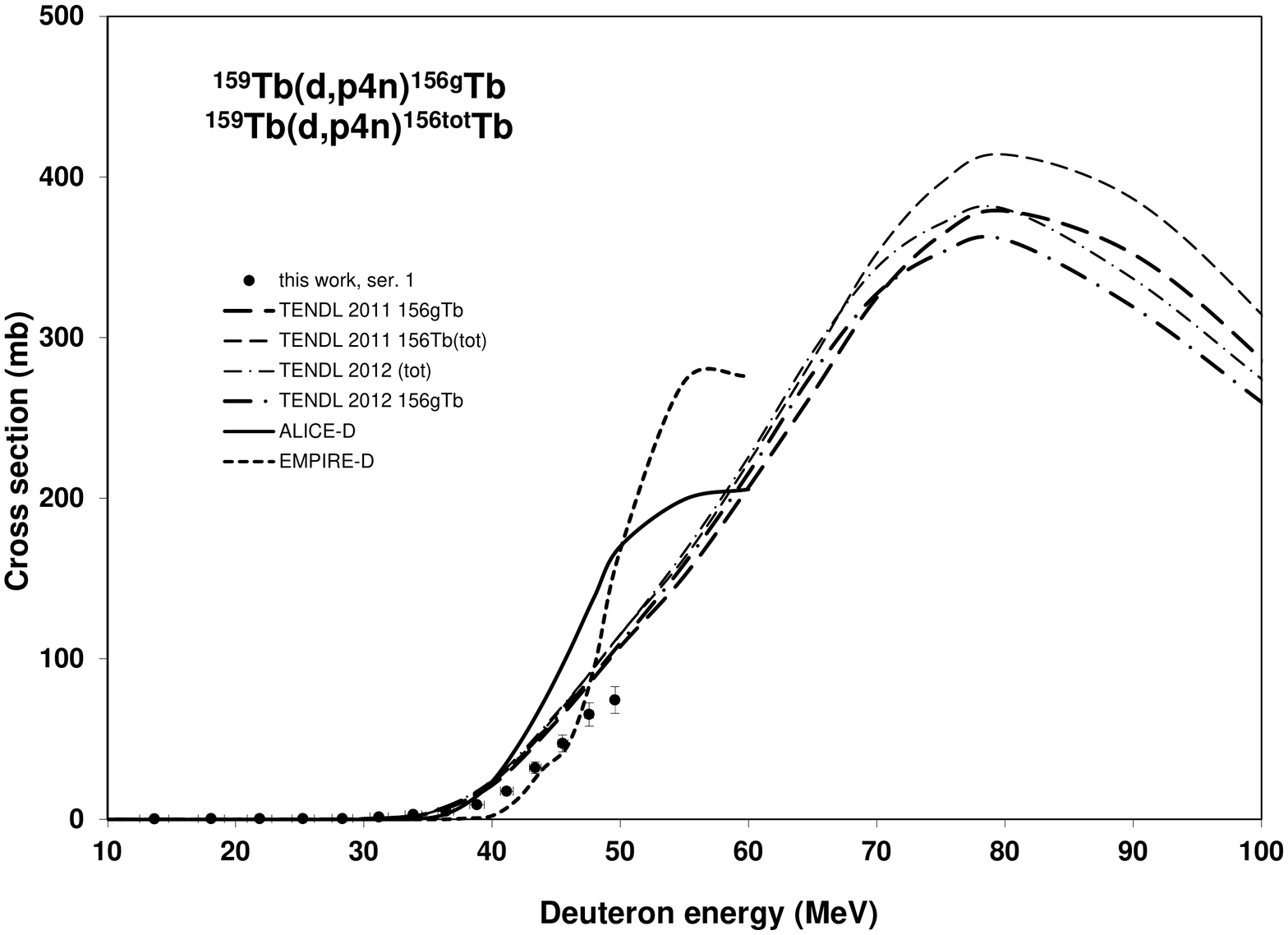}
\caption{Experimental and theoretical excitation functions of the $^{159}$Tb(d,p4n)$^{156g}$Tb(m+)  reaction}
\end{figure}

\subsubsection{Production of  $^{155}$Tb  (T$_{1/2}$=5.32 d) (cum)}
\label{4.1.6}
The measured experimental data (Fig. 7) are cumulative and are the sum of direct production and total decay of the shorter lived $^{155}$Dy (9.9 h) parent radioisotope. By referring to the excitation function of the $^{155}$Dy (Fig. 4 ) the main contributor to the cumulative cross-section of $^{155}$Tb is the decay of the parent $^{155}$Dy. This is confirmed by the results in the TENDL libraries. The best result is given by the EMPIRE-D model calculation, while ALICE-D and TENDL results overestimate the experimental values for the cumulative case.

\begin{figure}[h]
\includegraphics[scale=0.3]{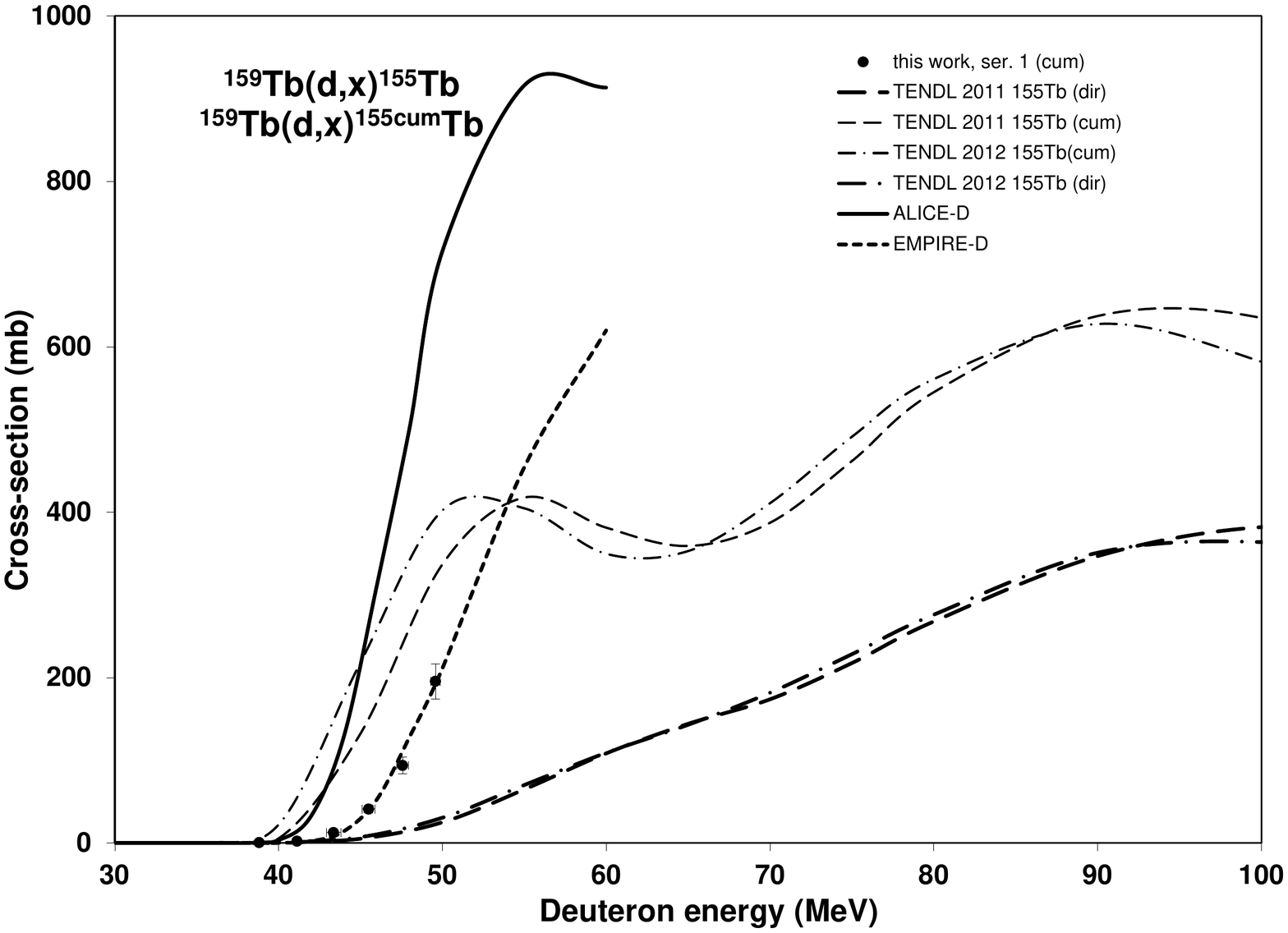}
\caption{Experimental and theoretical excitation functions of the $^{159}$Tb(d,x)$^{155}$Tb(cum)  reaction}
\end{figure}

\subsubsection{Production of $^{153}$Gd (T$_{1/2}$=240.4 d) (cum)}
\label{4.1.7}
The experimental data for cumulative production of the $^{153}$Gd are shown in Fig. 8. The experimental data were deduced from spectra measured after decay of the parent decay chain ($^{153}$Dy  6.4 h, $^{153}$Tb 2.34 d). No earlier experimental data are available. The theoretical calculations display a rather large spread of the corresponding yields. For all codes the main contributions below 60 MeV are connected with the $\alpha$-particle emission channels and the additional increases of the production cross-sections above 60 MeV are produced by the cumulative channels related to the decay of predecessors. TENDL-2012 underestimates distinctly the direct production of $^{153}$Gd, while ALICE-D and EMPIRE-D strongly overestimate it. The obtained differences of the code results are the consequence of models used for a simulation of the $\alpha$-particle emission. 

\begin{figure}[h]
\includegraphics[scale=0.3]{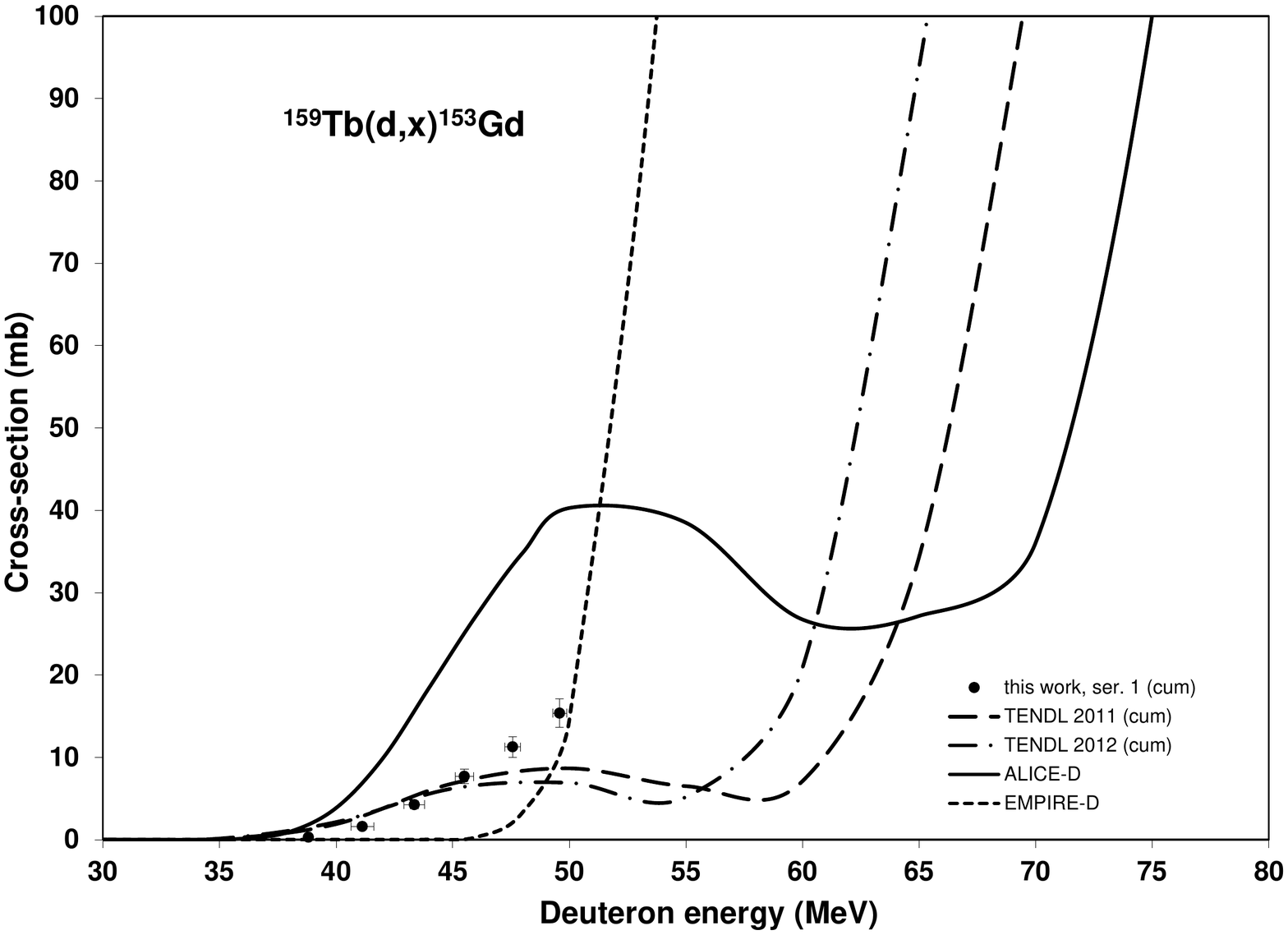}
\caption{Experimental and theoretical excitation functions of the $^{159}$Tb(d,x)$^{153}$Gd(cum)  reaction}
\end{figure}

\begin{table*}[ht]
\tiny
\caption{Measured cross-sections of the${}^{ }$${}^{159}$Tb(d,x)${}^{1}$${}^{59,157,155}$Dy, ${}^{159}$Tb(d,x)${}^{160,156g,155}$Tb  and  ${}^{159}$Tb(d,x)${}^{153}$Gd reactions}
\centering
\begin{center} 
\begin{tabular}{|p{0.3in}|p{0.2in}|p{0.3in}|p{0.2in}|p{0.3in}|p{0.2in}|p{0.3in}|p{0.2in}|p{0.3in}|p{0.2in}|p{0.3in}|p{0.2in}|p{0.3in}|p{0.2in}|p{0.3in}|p{0.2in}|} \hline 
\multicolumn{2}{|c|}{} & \multicolumn{14}{|c|}{ \textbf{Cross-section $\sigma$ $\pm$ $\Delta\sigma$ (mb)}} \\ \cline{3-16} 
\multicolumn{2}{|c|}{ \textbf{E $\pm\Delta$E (MeV)}} & \multicolumn{2}{|c|}{$^{159}$Dy} & \multicolumn{2}{|c|}{${}^{157}$Dy} & \multicolumn{2}{|c|}{${}^{155}$Dy} & \multicolumn{2}{|c|}{${}^{160}$Tb${}^{  }$} & \multicolumn{2}{|c|}{${}^{156g}$Tb} & \multicolumn{2}{|c|}{${}^{15}$${}^{5}$Tb} & \multicolumn{2}{|c|}{${}^{153}$Gd} \\ \hline 
 &  &  &  &  &  &  &  &  &  &  &  &  &  &  &  \\ \hline 
3.3 & 1.3 & 39.1 & 4.5 & 0.5 & 0.1 &  & ~ & 48.8 & 5.5 & 0.4 & 0.0 &  & ~ &  & ~ \\ \hline 
11.0 & 1.2 & 402.3 & 45.2 & 2.3 & 0.3 &  & ~ & 196.5 & 22.1 & 0.4 & 0.1 &  & ~ &  & ~ \\ \hline 
16.0 & 1.1 & 252.1 & 28.4 & 15.0 & 1.7 &  & ~ & 126.7 & 14.2 & 0.5 & 0.1 &  & ~ &  & ~ \\ \hline 
20.0 & 1.0 & 134.8 & 15.2 & 125.6 & 14.1 &  & ~ & 100.4 & 11.3 & 0.5 & 0.1 &  & ~ &  & ~ \\ \hline 
23.6 & 0.9 & 99.4 & 11.2 & 171.4 & 19.2 &  & ~ & 74.9 & 8.4 & 0.6 & 0.1 &  & ~ &  & ~ \\ \hline 
26.8 & 0.8 & 81.0 & 9.2 & 511.2 & 57.4 &  & ~ & 69.2 & 7.8 & 0.6 & 0.1 &  & ~ &  & ~ \\ \hline 
29.8 & 0.7 & 70.3 & 7.9 & 726.0 & 81.5 &  & ~ & 56.3 & 6.3 & 1.6 & 0.2 &  & ~ &  & ~ \\ \hline 
32.5 & 0.7 & 65.9 & 7.6 & 867.4 & 97.4 &  & ~ & 55.3 & 6.2 & 3.1 & 0.4 &  & ~ &  & ~ \\ \hline 
35.1 & 0.6 & 61.5 & 6.9 & 764.6 & 85.8 &  & ~ & 47.3 & 5.3 & 5.1 & 0.6 &  & ~ &  & ~ \\ \hline 
36.1 & 0.6 & 69.6 & 7.8 & 940.5 & 105.6 &  & ~ &  & ~ &  & ~ &  & ~ &  & ~ \\ \hline 
37.6 & 0.6 & 57.0 & 6.5 & 654.2 & 73.4 &  & ~ & 44.4 & 5.0 & 9.3 & 1.1 & 0.4 & 0.1 & 0.3 & 0.1 \\ \hline 
38.2 & 0.5 & 63.2 & 7.1 & 828.0 & 93.0 &  & ~ &  & ~ &  & ~ &  & ~ &  & ~ \\ \hline 
40.0 & 0.5 & 48.7 & 5.6 & 536.4 & 60.2 & 1.0 & 0.1 & 41.2 & 4.6 & 17.6 & 2.0 & 2.0 & 0.2 & 1.6 & 0.2 \\ \hline 
40.4 & 0.5 & 58.6 & 6.6 & 687.5 & 77.2 &  & ~ &  & ~ &  & ~ &  & ~ &  & ~ \\ \hline 
42.3 & 0.4 & 49.2 & 5.6 & 433.6 & 48.7 & 10.9 & 1.2 & 37.3 & 4.2 & 32.2 & 3.6 & 12.5 & 1.4 & 4.3 & 0.5 \\ \hline 
42.5 & 0.4 & 51.7 & 5.8 & 558.8 & 62.7 & 2.2 & 0.3 &  & ~ &  & ~ &  & ~ &  & ~ \\ \hline 
44.4 & 0.4 & 42.0 & 4.8 & 359.4 & 40.3 & 21.1 & 2.4 & 34.8 & 3.9 & 47.4 & 5.3 & 40.9 & 4.5 & 7.7 & 0.9 \\ \hline 
44.6 & 0.4 & 48.6 & 5.5 & 450.2 & 50.5 & 20.8 & 2.3 &  & ~ &  & ~ &  & ~ &  & ~ \\ \hline 
46.5 & 0.3 & 41.5 & 4.7 & 317.0 & 35.6 & 70.2 & 7.9 & 33.2 & 3.7 & 65.5 & 7.4 & 93.6 & 10.3 & 11.3 & 1.3 \\ \hline 
46.6 & 0.3 & 44.8 & 5.1 & 380.8 & 42.8 & 86.5 & 9.7 &  & ~ &  & ~ &  & ~ &  & ~ \\ \hline 
48.5 & 0.3 & 41.8 & 4.8 & 321.0 & 36.0 & 97.8 & 11.0 &  & ~ &  & ~ &  & ~ &  & ~ \\ \hline 
48.6 & 0.3 & 35.8 & 4.1 & 260.6 & 29.2 & 177.0 & 19.9 & 31.9 & 3.6 & 74.5 & 8.4 & 195.5 & 21.3 & 15.4 & 1.7 \\ \hline 
\end{tabular}
\end{center}
\end{table*}

\begin{table*}[ht]
\tiny
\caption{Production  routes of $^{159}$Dy and $^{157}$Dy}
\centering
\begin{center} 
\begin{tabular}{|p{0.7in}|p{0.5in}|p{1.0in}|p{0.8in}|p{0.5in}|p{0.8in}|} \hline 
\multicolumn{3}{|c|}{ ${}^{159}$Dy} & \multicolumn{3}{|c|}{${}^{157}$Dy} \\ \hline 
Reaction  & Energy range  & Ref & \textit{Reaction } & Energy range  & \textit{Reference} \\ \hline 
${}^{159}$Tb(p,n) & 25-6 &   \cite{8,33} & \textit{${}^{159}$Tb(p,3n)} & 30-20 &   \textit{\cite{7}} \\ \hline 
${}^{159}$Tb(d,2n) & 20-10 &   \cite{7, 10}, this work & \textit{${}^{159}$Tb(d,4n)} & 50-25 &   \textit{\cite{11}, this work} \\ \hline 
${}^{156}$Gd($\alpha $,n) & 25-15 &   TENDL & \textit{${}^{154}$Gd($\alpha $,n)} & 25-15 &   TENDL\textit{} \\ \hline 
${}^{157}$Gd $\alpha $,2n) & 30-18 &   TENDL & \textit{${}^{155}$Gd $\alpha $,2n)} & 30-20 &   TENDL\textit{} \\ \hline 
${}^{nat}$Gd($\alpha $,xn) & 50-20 &   TENDL & \textit{${}^{155}$Gd(${}^{3}$He,n)} & 30-15 &   TENDL\textit{} \\ \hline 
${}^{157}$Gd(${}^{3}$He,n) & 30-15 &   TENDL & \textit{${}^{156}$Gd(${}^{3}$He,2n)} & 50-20 &   TENDL\textit{} \\ \hline 
${}^{158}$Gd(${}^{3}$He,2n) & 40-18 &   TENDL &  &    & \textit{} \\ \hline 
${}^{nat}$Gd(${}^{3}$He,xn) & 50-20 &   TENDL &  &    &  \\ \hline 
\end{tabular}
\end{center}
\end{table*}

\subsection{Integral yields}
\label{4.2}
The so called physical integral yield (instantaneous short irradiation) \cite{22} as a function of the incident deuteron energy was calculated from a spline fit to our experimental data and are shown in Fig. 9 and Fig. 10. The only measurement find in the literature is from Mukhamedov \cite{12} for $^{160}$Tb and $^{159}$Dy. Our data are significantly higher than those given by Mukhamedov for both referred radioisotopes. 

\begin{figure}[h]
\includegraphics[scale=0.3]{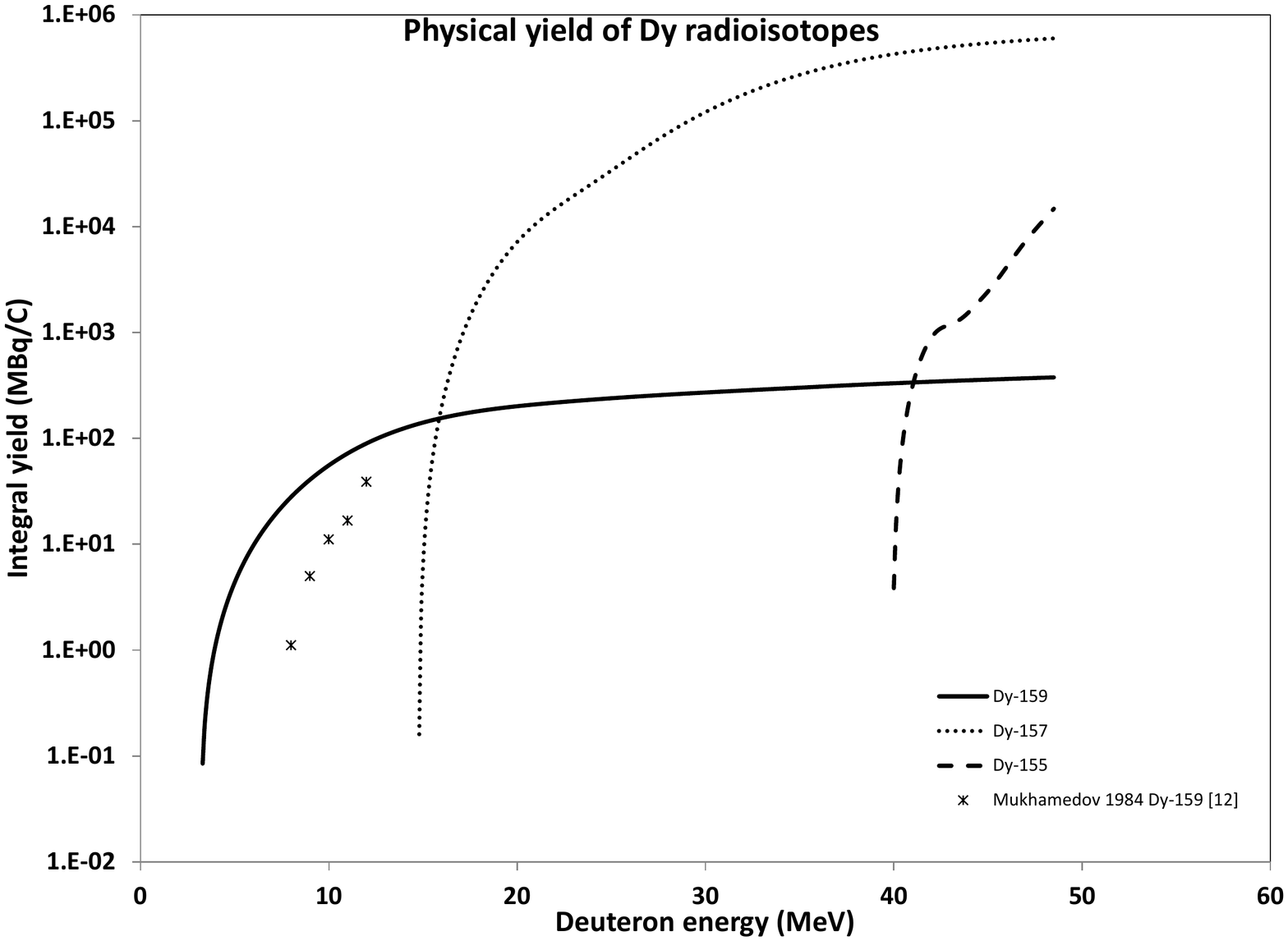}
\caption{Thick target yields for radionuclides of dysprosium produced by deuteron irradiation on $^{159}$Tb and comparison with the available literature values}
\end{figure}

\begin{figure}[h]
\includegraphics[scale=0.3]{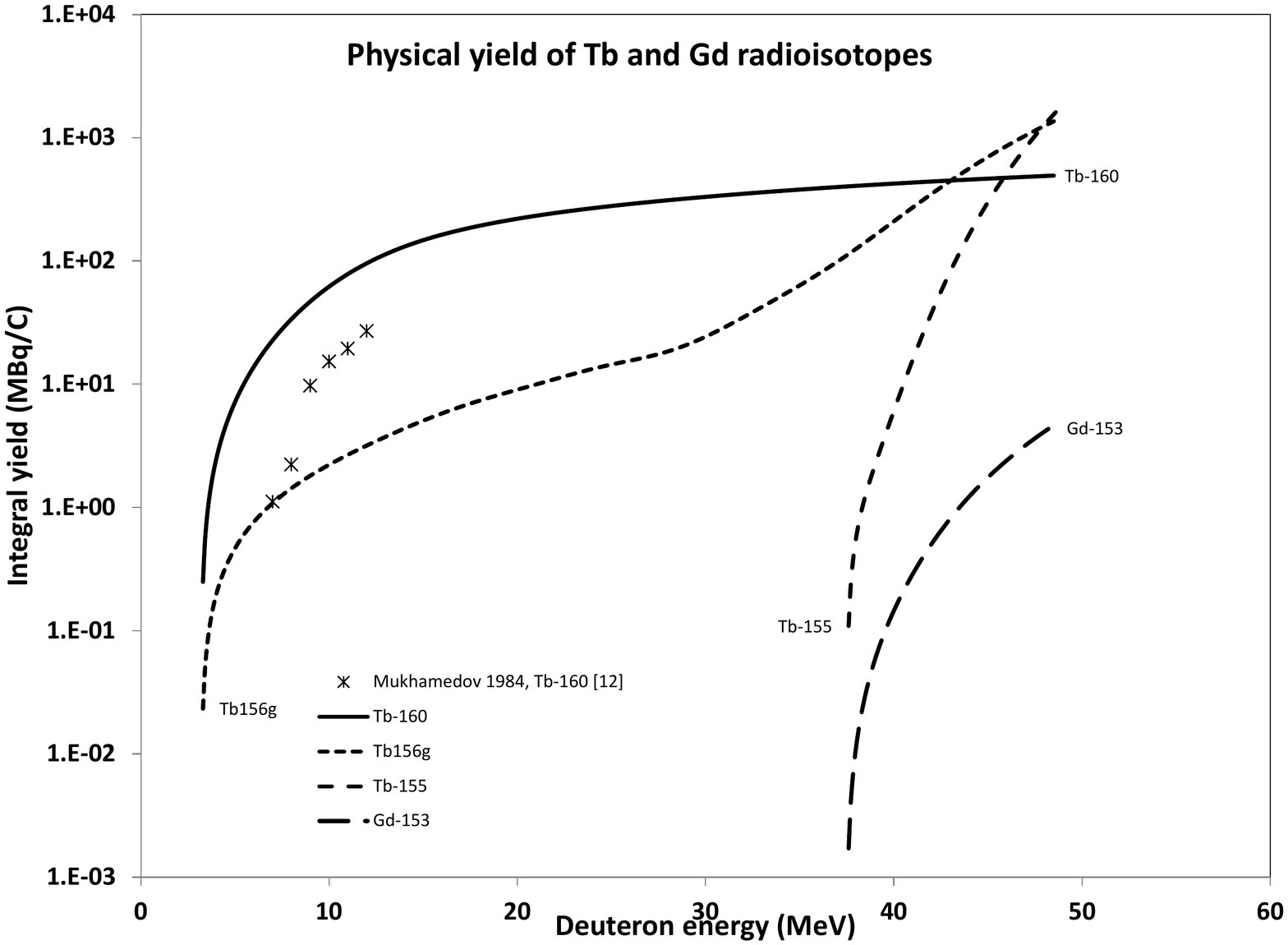}
\caption{Thick target yields for radionuclides of terbium and gadolinium  produced by deuteron irradiation on $^{159}$Tb and comparison with the available literature values}
\end{figure}

\section{Comparison of the experimental and theoretical results}
\label{5}
The comparison with the results of theoretical codes using global parameters shows that still large difference can be observed between the experimental results and the predictions the different codes. The new improvements for better description of the complex deuteron induced reactions resulted in a better description. The predictions for (d,pxn) reactions by of ALICE-D and EMPIRE-D are naturally the most successful, as  the correction included is based more directly on the systematics of experimental data. The agreement with the 2012 TENDL version is better, but the (d,p) reaction is still strongly under-predicted. 

\section{Comparison of the production routes of $^{157}$Dy and $^{159}$Dy}
\label{6}
Among the investigated reactions the routes leading to production of $^{157}$Dy and $^{159}$Dy have presently a practical interest. The two radioisotopes can be produced via various reactions. Out of these competitive production routes we discuss below the low energy, light charged particle methods in more detail, while summarizing briefly the other routes.

The $^{157}$Dy and $^{159}$Dy can be produced with the following bombarding beams and nuclear reactions:
\begin{itemize}
\item	By neutron induced reactions via (n,$\gamma$) on the neighboring stable Dy isotopes. The production yield is high. It requires highly enriched Dy targets to have radionuclidic pure end-products. The product is of low specific activity and is not carrier free \cite{30}.
\item	By photonuclear reactions \cite{31}. The product has usually low specific activity by using ($\gamma$,xn) at the presently used intensities and it is not carrier free.  The specific activity could be significantly increased by using the new high intensity  $\gamma$-beams. 
\item	By high energy spallation reaction and use of electromagnetic separators \cite{32}. The method assures a high specific activity, but the yield is low and the product will be expensive. It can be used in cases when no other method is available. Mostly in case of exotic and uncommon radioisotopes.
\item	By low and medium energy light charged particle induced nuclear reactions. The production yield is acceptable high. In favorable cases production does not require highly enriched targets. The end product is mostly carrier free and has high specific activity. 
\end{itemize}
For routine production, reactor production is often the most favorable, if the radionuclide fulfills the requirement of the applications (specific activity). Many medically interesting radionuclides can, however, be produced only with charged particle beams. The low and medium energy, high power (beams up to 1 mA) cyclotrons are also becoming more competitive for production of those radioisotopes, which presently produced in nuclear reactors. Photonuclear and spallation based methods are still not common, these methods need improvements on the beam intensity. The main production routes for $^{157}$Dy and $^{159}$Dy via light charged particle beam are summarized in Table 5 \cite{7,8,10,11,33}. The excitation functions of the reactions are shown in Fig. 11 and 12.

\begin{figure}[h]
\includegraphics[scale=0.3]{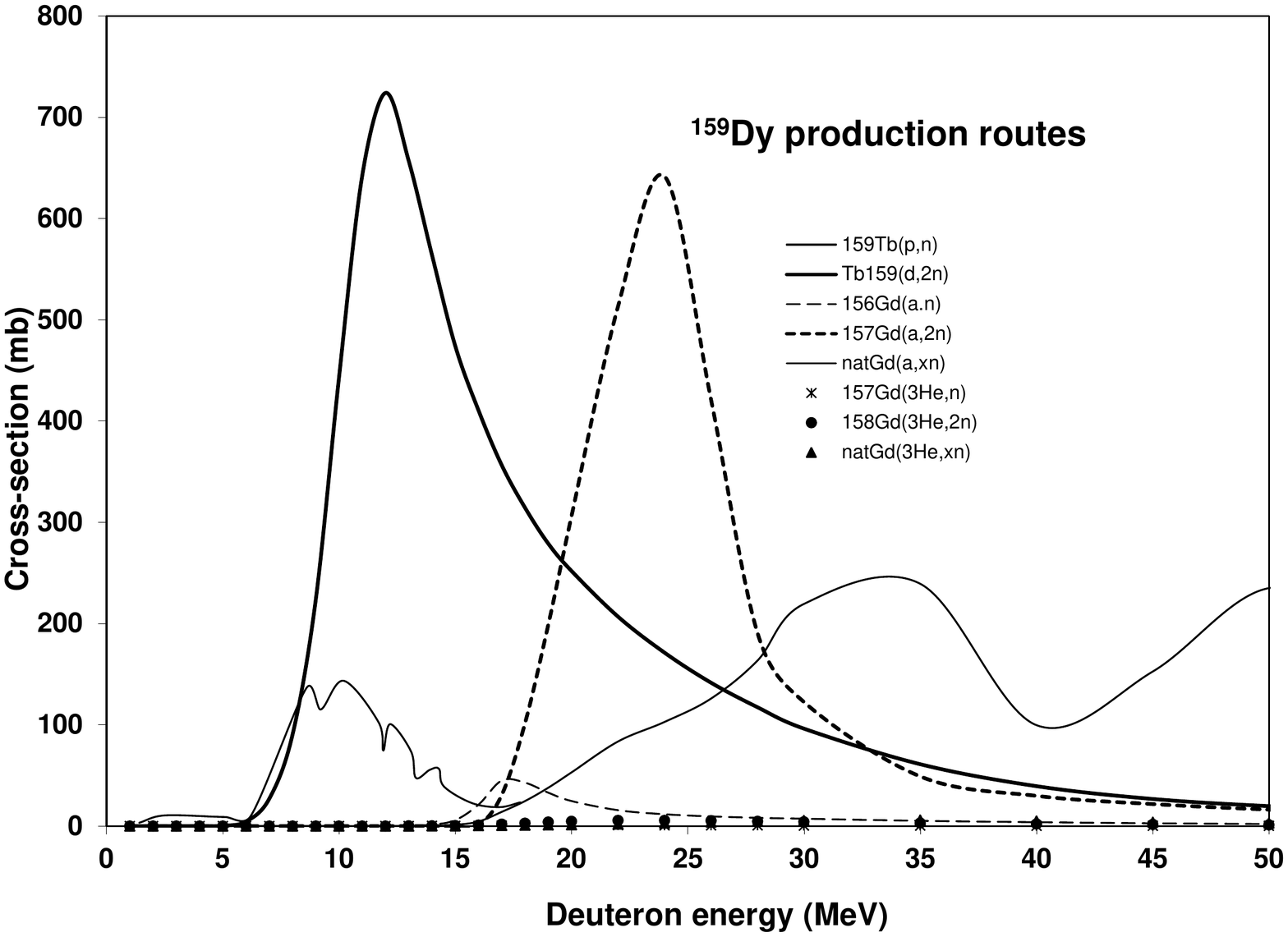}
\caption{Comparison of excitation functions for production of $^{159}$Dy}
\end{figure}

\begin{figure}[h]
\includegraphics[scale=0.3]{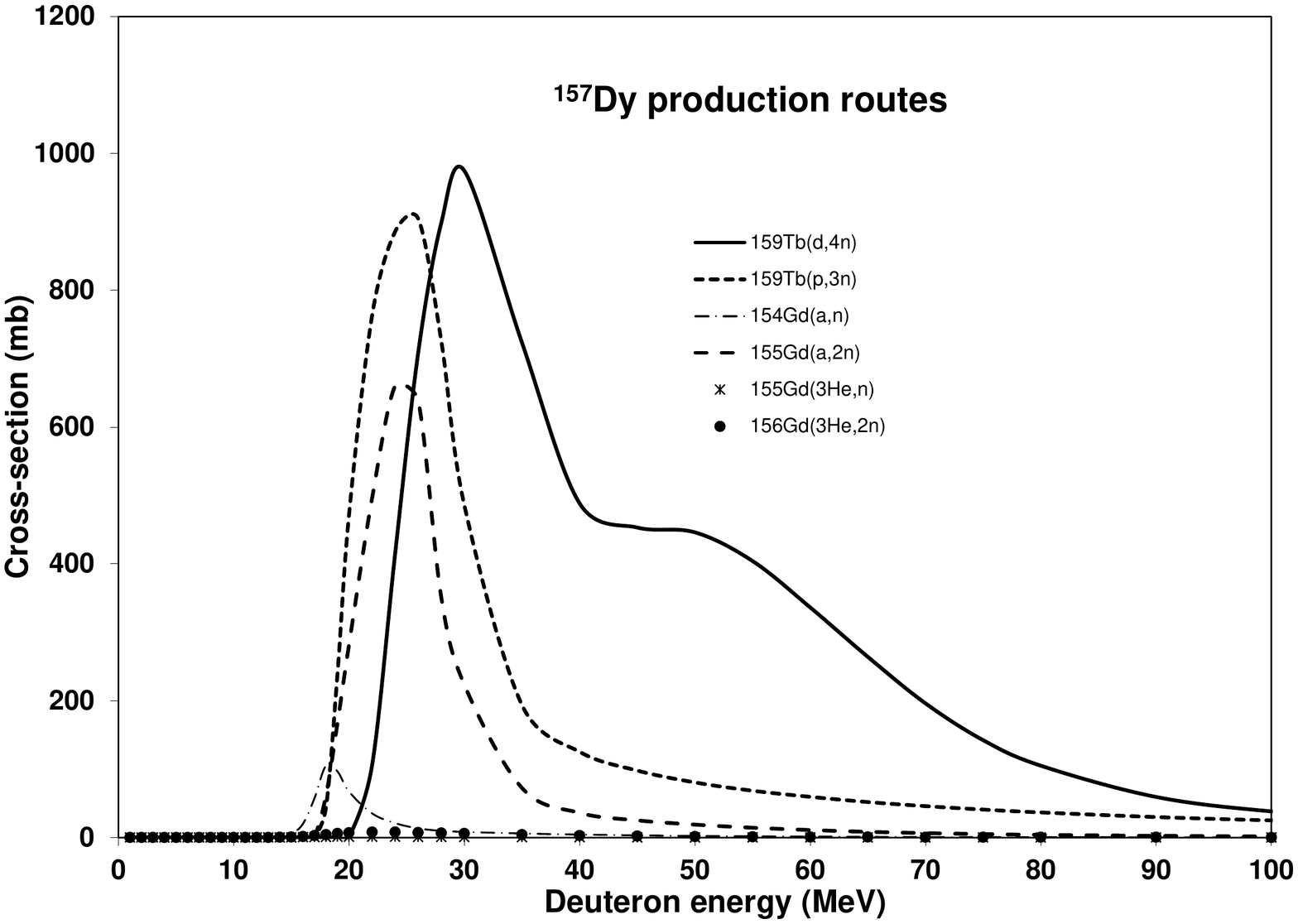}
\caption{Comparison of excitation functions for production of $^{157}$Dy}
\end{figure}

Taking into account that for many of the involved excitation function no experimental data exist, the excitation functions are based on the adjusted theoretical predictions of the TENDL 2012 library. The differences in the different routes are so high that conclusion is not affected by the known poor prediction capability of the theoretical results.

\subsection{Production routes of $^{159}$Dy}
\label{6.1}
According to Table 5 and Fig. 11 the most favorable route for production of $^{159}$Dy at a low energy cyclotron (k = 20) is the $^{159}$Tb(p,n) process and at higher energies (k = 35) the $^{159}$Tb(d,2n) reaction. The yield of the (d,2n) reaction is, like in many cases we discussed earlier, significantly higher than the yield of (p,n). According to Fig. 11 the cross-sections of the $^3$He induced reactions are very small, similar to cross-sections of the ($\alpha$,n) reaction.
The ($\alpha$,2n) process requires a 30 MeV cyclotron and highly enriched target. The cross-section is high but has lower yield comparing to the (d,2n) reaction due to the shorter range of the alpha particle.
The $^{159}$Dy can also be produced with high radionuclide purity on $^{nat}$Gd by both ($^3$He,xn) and alpha particle induced reactions. In both cases the yield is low.

\subsection{Production routes of $^{157}$Dy}
\label{6.2}
According to the Fig. 12 at a low energy cyclotron only the low yield $^{154}$Gd($\alpha$,n) reaction is available. At commercial 30-35 MeV H$^-$-cyclotrons the $^{159}$Tb(p,3n) reaction results in high production yields. In the optimum (30-20 MeV) energy range the amount of the simultaneously produced $^{159}$Dy is small due to the low cross-sections of the (p,n) in this higher energy range (see Fig. 12).
The energy range for the (d,4n) reaction should be 50-30 MeV to minimize the amount of $^{159}$Dy produced simultaneously via (d,2n). Accelerators capable of practical implementation of the (d,4n) reaction are however very rare. 
The $^{155}$Gd($\alpha$,2n) requires highly enriched targets and cyclotrons with 30 MeV alpha beams, and the yield is significantly lower compared to the $^{159}$Tb(p,3n) reaction.
To get high radionuclide purity highly enriched $^{155}$Gd and $^{156}$Gd targets should be used in case of $^3$He-particle induced reactions. The yield however is very low and the 3He beam is expensive even having recovery system for the gas in the ion source.

\section{Summary and conclusion}
\label{7}
The present work provides new data for the excitation function of deuteron induced nuclear reactions on natural terbium, which is actually $^{159}$Tb. The experimental results for the produced radioisotopes were compared with the former literature values as well as with the results of theoretical model calculations by using the codes ALICE-D, EMPIRE-D and TALYS 1.4 (the data taken from the TENDL 2011 and 2012 libraries). Our new measurements both verified or improved the former results and gave an extension to them. Comparison with the results of model codes could validate the recent improvements of those, as well as provide new basis for further improvements. As practical results, physical yield curves were also calculated for medical and other practival use. Serving for the present demands of radioisotopes research a detailed comparison is given for the possible routes for production of 2 important radioisotopes $^{157}$Dy and $^{159}$Dy.

\section{Acknowledgements}
\label{8}
This study was partly performed in the frame of the MTA-FWO (Vlaanderen) collaboration programs. The authors thank the different research projects and their respective institutions for the practical help and providing the use of the facilities for this study.
 



\clearpage
\bibliographystyle{elsarticle-num}
\bibliography{Tbd}







\end{document}